\documentclass[aps,prl,twocolumn,superscriptaddress,amsfont,amssymb,amsmath,nofootinbib,showpacs,balancelastpage]{revtex4-1}
\usepackage{float}
\usepackage{epsfig}
\usepackage{graphicx}
\usepackage{dcolumn}
\usepackage{bm}
\usepackage{color}
\usepackage{tabularx}
\usepackage{cleveref}

\def\thebiblio#1{
\begin{center}\bf \large References
\end{center}
\list
{[\arabic{enumi}]}{\settowidth\labelwidth{#1.}\leftmargin\labelwidth
 \advance\leftmargin\labelsep
 \usecounter{enumi}}
 \def\newblock{\hskip .11em plus .33em minus -.07em}
 \sloppy
 \sfcode`\.=1000\relax}

\begin{document}

\preprint{}
\title{ Isotropy theorem for arbitrary-spin cosmological fields
}

\author{J.A.R. Cembranos}
\email{cembra@ucm.es}
\author{A.L. Maroto}
\email{maroto@ucm.es}
\author{S.J. N\'u\~nez Jare\~no}
\email{sjnjareno@ucm.es}
\affiliation{Departamento de F\'{\i}sica Te\'orica, Universidad Complutense de Madrid, 28040 
Madrid, Spain}

\date{\today}

\begin{abstract}
We show that the energy-momentum tensor of homogeneous fields of arbitrary spin
in an expanding universe is always isotropic
in average provided the fields remain bounded and evolve rapidly
compared to the rate of expansion. An analytic expression for the
average equation of state is obtained for Lagrangians with
generic power-law kinetic and potential terms.  As an example we
consider the behavior of a  spin-two field in the  standard
Fierz-Pauli theory of  massive gravity. The results can be extended to general
space-time geometries for locally inertial observers.
\end{abstract}

\pacs{98.80.-k, 98.80.Cq}
\maketitle

\section{Introduction}

One of the main limitations on the use of  vectors or higher-spin fields
in cosmology is the high degree of isotropy of the universe on large scales  \cite{PlanckIsotropy}.
A homogeneous  field of non-zero spin generically breaks isotropy by selecting
preferred directions in  space.

However in recent years there has been a growing interest in the possibility
of using vectors fields (abelian or non-abelian) as dark matter \cite{Nelson}, dark energy  \cite{VT}
or inflaton candidates \cite{vectorinflation}. In these cases the anisotropy problem is avoided
thanks to the use of particular field configurations
(temporal components, triads, etc \cite{solution, Galtsov, Zhang})  that guarantee an isotropic
energy-momentum tensor.
Also a more general result has been proved   which shows that in the case
of bounded fields which evolve rapidly as compared to the rate of expansion,
the temporal average of the energy-momentum tensor is always isotropic
for any field configuration. This means, that even anisotropic field configurations
such as a linearly polarized field would give rise in average to an isotropic energy-momentum
tensor.  This result was obtained by using a generalization
of virial theorem and applies both to abelian \cite{Isotropy} and non-abelian  \cite{IsotropyYM} theories,
with arbitrary potentials  and with or without gauge-fixing terms.

The generality of this result for homogeneous vectors suggests
that the isotropy property could be a general feature of any field theory
for arbitrary spin with the only requirements of large scale homogeneity,
boundedness and rapid evolution. In this work we prove that this is indeed
the case and present a general isotropy theorem for arbitrary-spin cosmological
fields.


Unlike previous works in which explicit Lagrangian densities were used,
in the case of generic theories as those we will consider in this work, the explicit dependence of the Lagrangian
on the metric tensor is not fixed a priori. This means that we cannot use
the Hilbert form of the energy-momentum tensor:
\begin{eqnarray}
T^{\mu \nu} =-\frac{2}{\sqrt{g}}\frac{\delta S}{\delta g_{\mu\nu}}
\label{Hilbert}
\end{eqnarray}
 as our starting point.
In order to avoid this difficulty, we will make use of the so called
Belinfante-Rosenfeld  \cite{Belinfante, Rosenfeld} energy-momentum tensor, which allows to relate
the Hilbert energy-momentum with the canonical one by means of  the use
of some extra terms. Unexpectedly, we will show how this relation
between the canonical and Hilbert forms
is intimately related to the anisotropy issue.

For clarification, let us thus start by briefly reviewing the standard Belinfante-Rosenfeld
approach in Minkowski space-time  \cite{Itzykson} and consider a Lagrangian density depending only on the fields (labelled by $A$) and their gradients:
\begin{equation}
\mathcal{L} \equiv \mathcal{L}\left[ \phi^A, \partial_{\mu} \phi^A \right]\;,
\end{equation}
Under an infinitesimal $x$-dependent translation $x^\mu\rightarrow x^\mu+\delta a^\mu (x)$,
the field and its gradient change as \cite{Itzykson}:
\begin{eqnarray}
\delta \phi^A &=& \delta a^{\mu}(x)\; \partial_{\mu}\phi^A (x)\;,
\\
\delta \partial_{\mu} \phi^A (x) &=& \delta a^{\nu}(x) \; \partial_{\nu}\partial_{\mu} \phi^A(x) \nonumber
\\
&+& \partial_{\mu} \left[ \delta a^{\nu}(x) \right] \partial_{\nu}\phi^A(x)\;.\,\,\,
\end{eqnarray}
By imposing:
\begin{eqnarray}
0=\delta \int d^4x {\cal L} = - \int d^4 x \; \delta a_{\nu} \partial_{\mu} \Theta^{\mu \nu}\;,\label{perturbation}
\end{eqnarray}
we obtain that the canonical energy-momentum tensor, defined as
\begin{eqnarray}
\Theta^{ \mu \nu}= - \eta^{ \mu \nu} \mathcal{L} + \frac{\partial \mathcal{L}}{\partial \left( \partial_\mu \phi^A\right)} \partial^{\nu} \phi^A\;
\end{eqnarray}
is conserved:
\begin{eqnarray}
\partial_{\mu} \Theta^{\mu \nu} = 0\;.
\end{eqnarray}
This tensor is nothing but the Noether current associated to the symmetry under space-time translations.
Notice that although it is conserved,  $\Theta_{\mu\nu}$ is not necessarily symmetric.

However, this current is not unique, and we can add a new piece:
\begin{eqnarray}
\partial_{\rho} \tilde{\Theta}^{\rho\mu  \nu}\;,
\end{eqnarray}
with $ \tilde{\Theta}^{\rho\mu \nu}$ antisymmetric in the
first two indices. This new piece does not modify the value of the Noether charge because it is a total derivative, neither its time conservation because of its antisymmetry,
\begin{eqnarray}
Q ^{\nu}&=& \int d^3x (\Theta^{0 \nu}+\partial_{\rho} \tilde{\Theta}^{\rho 0  \nu} )\nonumber
\\
&=& \int d^3x (\Theta^{0 \nu}+\partial_{i} \tilde{\Theta}^{i 0  \nu} ) =\int d^3x \Theta^{0 \nu}\;;
\\
\frac{d Q^{\nu}}{d t}&=&\int d^3x (\partial_{\mu} \Theta^{\mu \nu}+\partial_{\mu}\partial_{\rho} \tilde{\Theta}^{\rho\mu  \nu} ) =0\;.
\end{eqnarray}
 We are interested in a symmetric energy-momentum tensor, i.e. that  required
to appear on the right hand side of Einstein equations. The new piece that must be added read \cite{EL}:
\begin{eqnarray}
T^{\mu \nu} = \Theta^{\mu \nu} - \frac{1}{2}\partial_{\rho}\left( S^{\rho \mu \nu} + S^{\mu \nu \rho} - S^{\nu \rho \mu} \right)\;,
\end{eqnarray}
with
\begin{eqnarray}
S^{\mu \nu \rho}= \Pi^{\mu}_A \Sigma^{\nu \rho} \phi^A\;,
\end{eqnarray}
where $\Sigma^{\nu \rho}$ are the antisymmetric  Lorentz group generators
in the corresponding representation and
\begin{eqnarray}
\Pi^{\mu}_A= \frac{\partial \mathcal{L}}{\partial \left( \partial_\mu \phi^A\right)}
\end{eqnarray}
is the generalized momentum associated to $\phi^A$.
$T^{\mu\nu}$ is the symmetric Belinfante-Rosenfeld energy-momentum tensor
which agrees with the Hilbert energy-momentum tensor obtained from
variations with respect to the metric (\ref{Hilbert}) as
shown in  \cite{Belinfante, Rosenfeld,Gri}.

Both, the canonical energy-momentum tensor $\Theta^{\mu\nu}$ and
the Belifante-Rosenfeld tensor $T^{\mu\nu}$ can be written in a
curved space-time in
a straightforward way by using minimal coupling, simply changing
ordinary derivatives by covariant ones, i.e.  we will work with:
\begin{eqnarray}
T^{\mu \nu} &=& \Theta^{\mu \nu}+\nabla_\rho \tilde{\Theta}^{\rho\mu \nu }
\nonumber \\&=& \Theta^{\mu \nu} - \frac{1}{2}\nabla_{\rho}\left( S^{\rho \mu \nu} + S^{\mu \nu \rho} - S^{\nu \rho \mu} \right)\;.
\end{eqnarray}
Notice, that the form of the
Lagrangian guarantees that only first derivatives of the fields
will appear in $\Theta_{\mu\nu}$.

\section{Homogeneous fields and virial theorem}

Following \cite{Isotropy} and \cite{IsotropyYM}, we can use a generalization of the virial theorem in order to obtain interesting results for the average energy-momentum tensor of 
homogeneous fields $\phi^A(t)$. Before writing the most general theorem, let us consider a Friedmann-Lema\^itre-Robertson-Walker (FLRW) metric for simplification:
\begin{eqnarray}
ds^2 = dt^2 - a(t)\;d\vec{x}^2\;.
\end{eqnarray}
With these assumptions, the $\tilde \Theta^{\rho\mu\nu}$ tensor is also homogeneous. 

Our aim is taking the temporal average of the energy momentum tensor during periods ${\cal T} \ll H^{-1}$, where $H$ is the Hubble parameter $H=\dot a /a$. Particularly, we are interested in the average value of $\nabla_\rho \tilde{\Theta}^{\rho\mu \nu }$ as this term will be the cause of the anisotropies.
\begin{eqnarray}
\left \langle \nabla_\rho\tilde{\Theta}^{\rho\mu \nu } \right \rangle =\frac{1}{{\cal T}} \int_{t}^{t+{\cal T}} dt' \left(\nabla_\rho\tilde{\Theta}^{\rho\mu \nu } \right)(t')\;, \label{virial1}
\end{eqnarray}
with
\begin{eqnarray}
 \nabla_\rho\tilde{\Theta}^{\rho\mu \nu }= \partial_0 \tilde{\Theta}^{0\mu \nu }
+ \left( \Gamma^\rho_{\delta \rho} \tilde{\Theta}^{\delta   \mu\nu } +  \Gamma^\mu_{\delta \rho} \tilde{\Theta}^{\rho\delta \nu } +  \Gamma^\nu_{\delta \rho} \tilde{\Theta}^{ \rho\mu\delta }  \right)\;. \nonumber\\
 \label{AnProm}
\end{eqnarray}
We can neglect the term in brackets on the right hand side of the equation (\ref{AnProm}) if the temporal derivative is larger than the expansion rate, i.e.  $\partial_0 \tilde{\Theta} \gg H \tilde{\Theta}$. If the system oscillates with an effective period $\tau$, $\partial_0 \Theta \sim \tau^{-1} \Theta$, then the condition for neglecting that term will be:
\begin{equation}
\tau^{-1} \gg {\cal T}^{-1} \gg  H\;.
\end{equation}

In this limit the energy-momentum tensor expressed in components reads
\begin{eqnarray}
T^{0 0}&=& \Pi^0_A \partial_0 \phi^A - \mathcal{L} -\frac{1}{2}\partial_0 \left( S^{0 0 0} \right)\nonumber \\
&=& \Pi^0_A\partial_0 \phi^A -\mathcal{L}\;;
\\
T^{0 j} &=& - \frac{1}{2} \partial_0 \left( S^{0 0 j}+S^{0 j 0}- S^{j 0 0}\right)= 0\;;
\\
T^{j j} &=& - g^{j j} \mathcal{L}-\frac{1}{2} \partial_0 \left( S^{0 j j} + S^{j j 0}-S^{j 0 j} \right)\nonumber
\\
&=& - g^{j j} \mathcal{L}-\partial_0 \left( \Pi^j_A \Sigma^{j 0} \phi^A \right)\;;
\\
T^{j k} &=& - \frac{1}{2} \partial_0 \left( \Pi^0_A \Sigma^{j k} \phi^A+ \Pi^j_A \Sigma^{k 0} \phi^A + \Pi^k_A \Sigma^{0 j} \phi^A \right)\;,\nonumber \\
\end{eqnarray}
with $k \neq j$. The antisymmetry of the Lorentz group generators, $\Sigma^{\mu \nu}$, has been used for simplification.

On the other hand, (\ref{virial1}) becomes
\begin{eqnarray}
\left \langle \nabla_\rho\tilde{\Theta}^{\rho\mu \nu }  \right \rangle &=& \frac{1}{{\cal T}} \int_{t}^{t+{\cal T}} dt'\partial_0 \tilde{\Theta}^{0\mu  \nu }(t') \nonumber
\\
&=& \frac{\tilde{\Theta}^{0\mu   \nu }(t+{\cal T})- \tilde{\Theta}^{0\mu   \nu }(t)}{{\cal T}}\;. \label{virial3}
\end{eqnarray}

As can be seen from (\ref{virial3}), if the field evolution is periodic or bounded, the right-hand side vanishes as  compared to $\langle T^{0 0} \rangle$ for sufficiently large ${\cal T}$. In fact, the ratio can be estimated as $\left \langle \nabla_\rho\tilde{\Theta}^{\rho\mu \nu }  \right \rangle/\langle T^{0 0} \rangle \sim \mathcal{O} \left( \tau / \cal{T}\right)$. That leads us to the following average energy-momentum tensor:
\begin{eqnarray}
\langle T^{0 0} \rangle &=& \langle \Pi^0_A \partial_0 \phi^A - \mathcal{L} \rangle\;;
\\
\langle T^{0 j} \rangle &=& T^{0 j} = 0\;;
\\
\langle T^{j j} \rangle &=& \langle - g^{j j} \mathcal{L} \rangle\;;
\\
\langle T^{j k} \rangle &=& 0\;; k \neq j\;,
\end{eqnarray}
which is explicitly isotropic.
Notice that as commented before,
 the anisotropies in the exact (non-averaged) tensor
indeed come from the new terms that
must be added in the Belinfante-Rosenfeld approach
in order to get the symmetric expression.

Moreover, using these results we can also express the average equation of state in this suggestive form:
\begin{eqnarray}
\omega = \frac{\langle p \rangle}{\langle \rho \rangle} = \frac{\langle \mathcal{L}\rangle}{\langle \Pi^0_A \partial_0 \phi^A - \mathcal{L}\rangle} = \frac{\langle \mathcal{L} \rangle}{\langle \mathcal{H} \rangle}\;, \label{omega1}
\end{eqnarray}
with $\mathcal{H}$ the Hamiltonian of the system.

There are other ways of writing this quantity:
\begin{eqnarray}
\omega = \frac{\langle \Pi^0_A \partial_0 \phi^A \rangle}{\langle \mathcal{H} \rangle} - 1\;.
\end{eqnarray}
Or by using the equation $\partial_0 \phi^A = \frac{\partial \mathcal{H}}{\partial \Pi^0_A}$:
\begin{eqnarray}
\omega= \frac{\langle \Pi^0_A \frac{\partial \mathcal{H}}{\partial \Pi^0_A} \rangle}{\langle \mathcal{H} \rangle}-1\;. \label{omega2}
\end{eqnarray}
Another form is reached by using the Euler-Lagrange equation for $\phi^A$, $\nabla_\mu \Pi^\mu_A= \frac{\partial \mathcal{L}}{\partial \phi^A}$:
\begin{eqnarray}
\omega = \frac{\langle \partial_0 \left( \Pi^0_A \phi^A \right)-\partial_0 \Pi^0_A \phi^A \rangle}{\langle \mathcal{H}\rangle}-1= \frac{\langle -\frac{\partial \mathcal{L}}{\partial \phi^A}\; \phi^A \rangle}{\langle \mathcal{H}\rangle}-1\;,
\nonumber \\ \label{omega3}
\end{eqnarray}
where we have also applied the extension of the virial theorem to $\Pi^0_A \phi^A$, i.e. $\langle \partial_0 \left(\Pi^0_A \phi^A \right) \rangle = 0$.

From (\ref{omega2}) and (\ref{omega3}), it can be seen that the following average equation is satisfied
\begin{eqnarray}
\langle \Pi^0_A \frac{\partial \mathcal{H}}{\partial \Pi^0_A} +\frac{\partial \mathcal{L}}{\partial \phi^A}\; \phi^A \rangle =0\;. \label{virial2}
\end{eqnarray}

 The last equation results very helpful when considering theories where the kinetic and  potential terms add separately as simple power-laws in
 the following form
 \begin{eqnarray}
\mathcal{H}= \left(\lambda^{AB} g_{00} \Pi^0_A \Pi^0_B\right)^{n_T} + \left( M_{AB} \phi^{A} \phi^{B} \right)^{n_V}\;,
 \end{eqnarray}
 where $\lambda^{A B}$ and $M_{A B}$ are constant matrices. In such a case, Equation (\ref{virial2}) relates $T$ and $V$
 in the following form
\begin{eqnarray}
\langle T \rangle = \frac{n_V}{n_T}\; \langle V \rangle\;.
\end{eqnarray}
By using (\ref{omega3}), we can obtain an analytic expression for $\omega$ independent of initial conditions or
particular polarization of $\phi^{A}$:
\begin{eqnarray}
\omega= \frac{2 \; n_V \langle V \rangle }{\langle T +V \rangle}-1= \frac{2\;n_V}{1+\frac{n_V}{n_T}}-1\;.
\end{eqnarray}
Notice that this result is also independent of the field spin. For instance, for the usual case with $n_T = 1$,
the behaviour of the equation of state is the same as that for scalar \cite{Turner} or vector \cite{Isotropy, IsotropyYM}
fields:
\begin{eqnarray}
\omega = \frac{n_V - 1}{n_V + 1}\;.
\end{eqnarray}

Note that fast oscillating fields can have associated a negative effective equation of state parameter.
In this sense, they are potential new models of
dark energy or inflation. Indeed, we have shown that this result does not depend on the spin. Similar 
approaches for scalar fields have been already considered in the literature \cite{RapidOscInflation,ThreeFormInflation}. Another potential interest of these results comes from the possibility of avoiding the anisotropy typically expected during the reheating period in inflationary models based on vectors or higher-spin fields.  

\section{A spin-2 example}

As  an example, we will apply the previous results to the Fierz-Pauli theory of massive gravity on a curved space-time background given by the Lagrangian
\footnote{Note that we are assuming a minimal gravitational coupling for the spin-2 field.
There are more general options \cite{Dietrich} but
they are not relevant for the isotropy theorem
presented in this analysis.}
\begin{eqnarray}
\mathcal{L} &=& \frac{M_{Pl}^2}{8} \Bigl[ \nabla_\alpha h^{\mu \nu} \nabla^\alpha h_{\mu \nu}
- 2 \nabla_{\alpha} h^\alpha_\mu \nabla_\beta h^{\mu \beta} \Bigl.
\nonumber
\\
&+& 2 \nabla_{\alpha} h^\alpha_\mu \nabla^\mu h^{\beta}_{\beta}
- \nabla_{\alpha} h^\mu_\mu \nabla^\alpha h^{\nu}_{ \nu}
\nonumber
\\
&-&\Bigl. m_g^2 \left( h_{\mu \nu}h^{\mu \nu} - \left(h^\mu_\mu \right)^2 \right) \Bigr]\;.
\end{eqnarray}
The momentum of this field can be written as
\begin{eqnarray}
\Pi_{\mu \nu}^0 &=& \frac{\partial \mathcal{L}}{\partial \left( \nabla_0 h^{\mu \nu}\right)}
= \frac{M_{Pl}^2}{4} \left[ \nabla^0 h_{\mu \nu} - 2 \delta_{(\mu}^0 \nabla_\alpha h_{\nu)}^\alpha\right.
\nonumber \\
&+& \left. \delta_{(\mu}^0 \nabla_{\nu)} h^\alpha_\alpha
+ g_{\mu \nu} \nabla_\alpha h^{\alpha 0} -  g_{\mu \nu} \nabla^0 h^\alpha_\alpha\; \right]\;,
\end{eqnarray}
where $A^{(\mu} B^{\nu)}= (A^\mu B^\nu+A^\nu B^\mu)/2$.

Imposing homogeneity, considering a FLRW metric and exploiting the fact that $h_{\mu \nu}$ is symmetric, the momenta and the Lagrangian take the form
\begin{eqnarray}
\Pi_{0 \mu}^0 &=& 0\;;\nonumber \\
\Pi_{i j}^0 &=& \frac{M_{Pl}^2}{4} \partial^0 h_{i j}\;,\;\; i\neq j\;;\nonumber \\
\Pi_{i i}^0 &=& - \frac{M_{Pl}^2}{4} \sum\limits_{j \neq i} \partial^0 h_{j j}\;;\;\;
\\
\mathcal{L} &=& \frac{M_{Pl}^2}{8} \left[ \partial_0 h_{i j} \partial^0 h^{i j} - \partial_0 h^i_i \partial^0 h^j_j \right.
\\
&-&\left. m_g^2 \left(h_{\mu \nu}h^{\mu \nu} - \left( h^{\mu}_\mu \right)^2 \right)\right]\;, \nonumber
\end{eqnarray}
where we have neglected the expansion rate with respect to the temporal variation of the field.
We will also need the explicit expression for the Hamiltonian. Under the same assumptions, we
can write
\begin{eqnarray}
\mathcal{H} &\equiv& \Pi_{\mu \nu}^0 \partial_0 h^{\mu \nu} - \mathcal{L}
= \frac{M_{Pl}^2}{8} \Bigl[ \partial_0 h_{i j} \partial^0 h^{i j}\Bigr.
\\
&-&\Bigl.
\partial_0 h^i_i \partial^0 h^j_j
+ m_g^2 \left(h_{\mu \nu}h^{\mu \nu} - \left( h^{\mu}_\mu \right)^2 \right)\Bigr]\;.\;\;\; \nonumber
\end{eqnarray}
As it can be seen, the Lagrangian and the Hamiltonian take the classical structure $\mathcal{L} = T - V$ and $\mathcal{H} = T + V$. If the field evolves under the conditions for applying the virial theorem, then (\ref{virial2}) holds. Consequently,
\begin{eqnarray}
& & \left\langle \Pi^0_{\mu \nu} \frac{\partial \mathcal{H}}{\partial \Pi^{0}_{\mu \nu}}
+\frac{\partial \mathcal{L}}{\partial h^{\mu \nu}}\; h^{\mu \nu} \right\rangle
\nonumber
\\
& & =
\left \langle \Pi^0_{\mu \nu} \nabla_0 h^{\mu \nu}
+\frac{\partial \mathcal{L}}{\partial h^{\mu \nu}}\; h^{\mu \nu}\right\rangle
\nonumber
\\
& & =
\left \langle 2 T - 2 V \right\rangle =0\;,
\end{eqnarray}
where one of the Hamilton equations has been used in the first equality. We can conclude that the behaviour of the field will be that of non-relativistic matter by using the last average equation and (\ref{omega1}):
\begin{eqnarray}
\omega = \frac{\left\langle \mathcal{L}\right\rangle}{\left\langle \mathcal{H}\right\rangle} = \frac{\left\langle T - V \right\rangle}{\left\langle T + V \right\rangle} = 0\;.
\end{eqnarray}
Therefore, given the weak coupling to matter fields, a homogeneous spin-two massive
graviton can contribute to the dark matter density. The massive graviton has been already studied as a dark matter candidate by assuming an isotropic stochastic background \cite{Dubovsky:2004ud,Pshirkov:2008nr}. However, even an anisotropic coherent 
evolution could  be taken into account as a viable model since, as shown before, 
it does not introduce an important amount of anisotropy in the background geometry.

\section{General geometrical backgrounds and discussion}

Finally, let us extend this result to a more general space-time
geometry by considering an inertial observer located at $x_0^\mu = 0$ and write the metric around it using Riemann normal coordinates:
\begin{eqnarray}
g_{\mu\nu}(x)=\eta_{\mu\nu}+\frac{1}{3}R_{\mu\alpha\nu\beta}x^\alpha x^\beta +\dots
\label{normal}
\end{eqnarray}
If the following conditions hold:
\begin{enumerate}
\item { The Lagrangian depends only on the fields and their gradients.}

\item {The field evolves rapidly:
\begin{eqnarray}
&&|R^\gamma_{\lambda\mu\nu}| \ll (\omega_{A})^{2} ,\;
\text{and} \;\;
|\partial_j S^{\mu \nu \rho}| \ll | \partial_0{S}^{\mu \nu \rho} | ,\;\;
\nonumber \\
&&\;\;\;\;\;\;\;\;\text{for}\;\; j=1,2,3\;;
\end{eqnarray}
for  any component of the Riemann tensor.  $\omega_{A}$ is the characteristic frequency of $\phi^A$.}

\item{ $S^{\mu \nu \rho}$, i.e. $\phi^{A}$ and $\Pi^{0}_A$, remains bounded in the evolution.}
\end{enumerate}
then, the second condition implies that if the averaging times satisfy
\begin{eqnarray}
|R^\gamma_{\lambda\mu\nu}| \ll {\cal T}^{-2} \ll (\omega_A)^{2}\;,
\end{eqnarray}
we are in a normal neighborhood and we can neglect the second
term in (\ref{normal}) so that we can  work locally in  a Minkowskian
space-time. In the normal neighborhood of
the observer, $\tilde \Theta^{\rho\mu\nu}$ can also
be considered as a homogeneous field. In such a region,  it is then possible to
rewrite all the above equations in Minkowski space-time ($a(t)=1$).
Accordingly,
it is possible to neglect the right-hand side in (\ref{virial3}) and prove that the mean value of the energy-momentum tensor
is isotropic. Thus, if
oscillations are fast compared to the curvature scale, the
average energy-momentum tensor takes the perfect fluid form for any locally inertial observer.

\vspace{0.2cm}

{\bf Acknowledgements}
This work has been supported by MICINN (Spain) project numbers FIS2011-23000, FPA2011-27853-01 and Consolider-Ingenio MULTIDARK CSD2009-00064.

\end{document}